# Effect of crystal field and external magnetic field on the magnetic and thermodynamic properties of the Ising mixed spin (1-1/2-1) three layers system of cubic structure.

Maen Gharaibeh[1], Ahmad Al-Qawasmeh[2], Abdalla Obeidat[1] and Sufian Abedrabbo[2,3]

[1]Department of Physics, Jordan University of Science and Technology, Irbid, Jordan
[2]Department of Physics, Khalifa University of Science and Technology, Abu Dhabi, UAE
[3]Department of Physics, University of Jordan, Amman, Jordan

*Corresponding author: magh@just.edu.jo

**Abstract**

In this work, Monte Carlo simulation was implemented to study the phase diagrams and magnetic properties of a mixed spin (1-1/2-1) three-layer system with a cubic structure. In particular, the effects of the exchange and coupling parameters on the system compensation temperature were explored. More importantly, the effect of the crystal field on the system compensation, critical temperatures, and magnetic and thermodynamic properties of the system have been explored in detail. The crystal field was found to have a profound impact on the system compensation and critical temperatures. Our results suggest that the presence of the crystal field will cause the compensation temperature to exist for a short range of negative crystal field values. Furthermore, the threshold values of the crystal field were found to determine the change in both the compensation and critical temperatures with the variation of the crystal field. The double-peak phenomena confirmed these results in the susceptibility and specific heat curves. Finally, the system magnetic hysteresis loops were studied for different temperatures and crystal field values, revealing the presence of two-step hysteresis loops for several values of these parameters.

## 1. Introduction

Ferromagnetic materials have recently been a subject of interest in experimental and theoretical physics owing to their wide range of applications in modern technology such as drug delivery [1], data storage devices [2], spintronic devices [3], sensors [4], and Li-ion batteries [5]. Experimental studies commonly emphasize the synthesis of materials using techniques such as atomic layer deposition [6-8] and characterization of their structural properties using techniques such as transmission electron microscopy [9]. On the other hand, theoretical studies have implemented different methods such as the effective field theory (EFT) [10-12] and Monte Carlo simulations [13-15] to study the magnetic and thermodynamic properties of these materials. In particular, the compensation temperature of ferromagnetic materials is a very important property to study from a theoretical perspective because its presence is of great technological significance. The compensation temperature is defined as a point below the critical temperature at which the sublattice magnetizations cancel out exactly. At this point, a small magnetic field is required to change the sign of the magnetization. Furthermore, magnetic hysteresis loops are very important for theoretical studies because the presence of multiple-step loops in the material is important for multistate memory devices [16].

Ferrimagnetic mixed spin Ising systems are simple yet reliable systems that have been widely used to study ferrimagnets. Despite its simplicity, the Ising model remains one of the most reliable theoretical models for simulating the magnetic properties of mixed spin structures. The Ising model can be solved exactly in special cases [17, 18], or it can be solved in the framework of several tools such as Monte Carlo simulations [19, 20]. Recently, the Monte Carlo approach has gained significant interest. The magnetic properties of several two-dimensional mixed spin structures such as cubic [21], triangular [22], honeycomb [23], and several shapes of the nanotube structure [13, 24, 25] were successfully simulated using this approach.

In our recent work [21], the Monte Carlo simulation approach was used to study the effect of exchange interaction on the critical temperature, $T_c$, and the compensation temperature, $T_{comp}$, for a mixed spin (1-1/2-1) three-layer system with a cubic structure. In that work, the systems $T_c$ and $T_{comp}$ were found to be sensitive to the exchange interaction constants between atoms with the same spin and are not sensitive to the exchange interaction between atoms with different spin. Furthermore, the system $T_{comp}$ was found to be less sensitive than $T_c$ to the variation in the number of layers of the system.

This work is motivated by several recent studies [10, 26-32], which found that the single-ion anisotropy has a profound impact on the magnetization behavior, critical and compensation temperatures, and magnetic hysteresis loops of different mixed spin structures.

In this work, we use the Ising model in the framework of Monte Carlo simulations to investigate the effect of the crystal field on the magnetization behavior in general and the critical and compensation temperatures, particularly for the mixed spin (1-1/2-1) three-layer system of cubic structure. Furthermore, the magnetic hysteresis loops for the system were analyzed in detail.

## 2. Model and formalism

*2.1 Lattice structure and Hamiltonian*

The triple-layer cubic structure studied in this study is shown in Fig. 1. The structure consists of two sublattices, A and B, ordered as ABA along the z-direction, forming a cube with side $L$ with the interface located along the $xy$-plane. Sublattice A forms two outer rectangular cuboids, each consisting of $L_A$ layers along the z-direction, and each layer consists of $N \times N$ $S$-type spin -1 atoms. Sublattice B is placed between the two outer A sublattices, and it consists of $L_B$ layers along the z-direction, and each layer consists of $N \times N$ $\sigma$-type spin -1/2 atoms. The total length of the cube layers is $L = 2L_A + L_B$ and the number of atoms in the structure was $N_{tot} = N^2 \times L$. The Hamiltonian terms consist of $S$-$S$, $\sigma$-$\sigma$ ferromagnetic interaction, and $S$-$\sigma$ ferrimagnetic interaction, as well as the interaction of spins with external magnetic and the interaction of $S$-type spins with the crystal field. The Hamiltonian of the system is given by:

$$H = -J_1 \sum_{<i,j>} S_i S_j - J_2 \sum_{<l,k>} \sigma_i \sigma_j - J_{12} \sum_{<i,l>} S_i \sigma_l - B \sum_i (S_i + \sigma_i) - D \sum_i S_i^2 \qquad (1)$$

where $J_1$, $J_2$ and $J_{12}$ are the coupling constants between spins $S$-$S$, $\sigma$-$\sigma$, and $S$-$\sigma$ respectively. The summation indices $<i,j>$ and $<l,k>$ denote the summations over the nearest-neighbor spins $S$-$S$ and $\sigma$-$\sigma$ respectively. The fourth term in the Hamiltonian represents the interaction between the external magnetic field $B$ and the $S$-type-and $\sigma$-type spins. The last term in the Hamiltonian represents the interaction of the crystal field with the $S$-type spin. In our simulations, we assign $J_1 > 0, J_2 > 0$ to ensure a ferromagnetic interaction of spins of the same type, and we selected $J_{12} < 0$ to ensure an antiferromagnetic interaction between spins of different types.

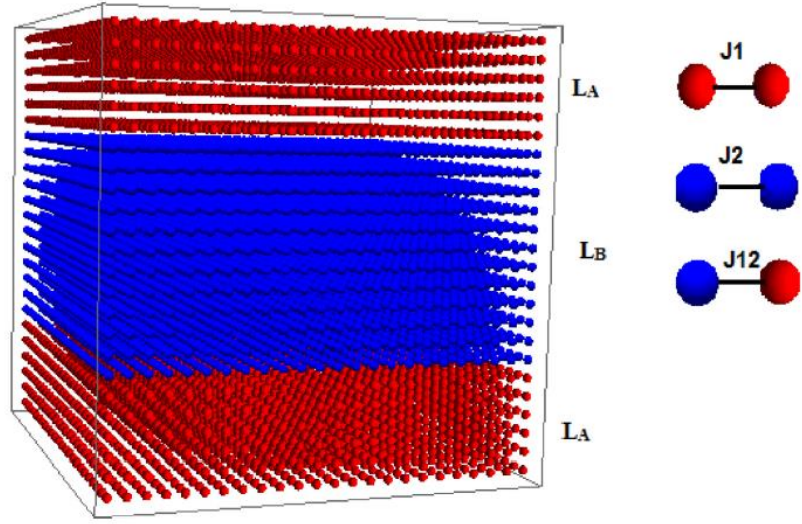

**Fig. 1.** Schematic representation of the triple layers cubic lattice. The outer layers colored red represent atoms of sublattice A with spin ($S = 0, \pm1$). The middle layers colored blue represents atoms of sublattice B with spin ($\sigma = \pm1/2$).

*2.2 Monte Carlo Simulation and Calculations*

The triple layer cubic structure shown in Fig. 1 was simulated using the Monte Carlo simulation method based on the metropolis algorithm [33]. Periodic boundary conditions were applied along the $x, y, z$. In this method, new configurations were generated by choosing a random spin of the $S$ or $\sigma$ types and flipping it randomly to the other possible values of the spin. Each spin is then accepted or rejected based on the Metropolis algorithm. A 100000 Monte Carlo step was used to equilibrate the system, followed by 100000 steps for each spin configuration. The results are reported for the system shown in Fig. 1, with a total number of atoms of 38400 correspondings to $L = 24$ (i.e., $2L_A = 12$, $L_B = 12$) and $N = 24$ atoms. For the error calculation, we use the method of blocks in which the $L$-size is divided into $n_b$ blocks of length $L_b = L/n_b$. The number of blocks is chosen such that $L_b$ is higher than the correlation length. Therefore, error bars were calculated by grouping all blocks and taking the standard deviation [34]. The magnetic properties of the system were calculated as follows:

The system total magnetization $M$, the magnetization $M_A$ for sublattice A and $M_B$ for sublattice B were calculated using the following equations:

$$M = \frac{M_A + M_B}{2} \tag{2}$$

$$M_A = \frac{\langle \sum_i S_i \rangle}{2N^2 L_A} \tag{3}$$

$$M_B = \frac{\langle \sum_i \sigma_i \rangle}{N^2 L_B} \tag{4}$$

The total magnetic susceptibility per site $\chi_{tot}$, sublattice A susceptibility $\chi_A$ and sublattice B susceptibility $\chi_B$ are given by the following equations:

$$\chi_{tot} = \frac{\chi_A + \chi_B}{2} \tag{5}$$

$$\chi_A = 2\beta \times N^2 \times L_A \left( \langle M_A^2 \rangle - \langle M_A \rangle^2 \right) \tag{6}$$

$$\chi_B = \beta \times N^2 \times L_B \left( \langle M_B^2 \rangle - \langle M_B \rangle^2 \right) \tag{7}$$

here $\beta = \frac{1}{k_B T}$ where $k_B$ is Boltzmann's constant, which is taken as 1 for simplicity. Finally, the system-specific heat $C_v$ is calculated as follows:

$$C_v = \frac{\beta^2}{N_{tot}} \left( \langle H^2 \rangle - \langle H \rangle^2 \right) \tag{8}$$

The compensation temperature is when the magnetization of two sublattices adds up to zero, given a zero-total magnetization of the lattice, as shown in Eq. (2). Therefore, the value of $T_{comp}$ is determined from the magnetization curves by locating the crossing point of the two sublattice magnetizations under the conditions given in Eq. (9) and Eq. (10):

$$|M_A(T_{comp})| = |M_B(T_{comp})| \tag{9}$$

$$sign(M_A(T_{comp})) = -sign\left(M_B(T_{comp})\right) \tag{10}$$

with $T_{comp} < T_c$, where $T_c$ is the critical temperature which is determined from the divergence of the susceptibilities curves. Eq. (9) and Eq. (10) indicate that the absolute values of the sublattice magnetizations are equal to each other; however, the sign of magnetization is different at the compensation point, $T_{comp}$.

## 3. Results and discussions

This section presents our results for the magnetic and thermodynamic properties of the mixed spin triple-layer cubic structure. In particular, the effect of the crystal field $D$ on these properties is thoroughly explored. Furthermore, the magnetic hysteresis loops for the system were analyzed in detail.

### 3.1 *Magnetic properties and phase diagrams*

In this section, the magnetic and thermodynamic properties of the triple-layer cubic structure, such as the total magnetization in Eq. (2), the sublattice magnetization in Eq. (3) and Eq. (4), respectively, and the total magnetic susceptibility per site in Eq. (5) and the specific heat in Eq. (8) will be presented at different values of the crystal field $D$. Furthermore, to explore the effect of $D$ on the system compensation temperature $T_{comp}$ defined in Eq. (9) and Eq. (10) and the critical temperature $T_c$ defined as the second zero of the magnetization curve, phase diagrams of $T_{comp}$ and $T_c$ will also be presented.

In our previous work [21] on this system, the effects of the coupling constants $J_1$, $J_2$ and $J_{12}$ on the system critical and compensation temperatures were explored. Both of $T_c$ and $T_{comp}$ were found to be sensitive to $J_1$ and $J_2$ and not to $J_{12}$. In particular, the compensation temperature $T_{comp}$ was found to be absent for particular values of $J_1$ and $J_2$.

For our calculations, it is essential to explore the effect of coupling constants $J_1$ and $J_2$ on the system compensation temperature more thoroughly by considering a wider range for the values of $J_1$ and $J_2$. Our calculated value for the system compensation temperature $T_{comp}$ at different values of $J_1$ and $J_2$ is shown in Fig. 2. Here, $J_{12}$ is fixed at -0.3, $D$ is fixed at zero, $J_1$ is varied from 0.1 to 0.8 in 0.1, and $J_2$ is varied from 0.5 to 2.5 in 0.25 steps. As shown in the figure, the system exhibits a compensation behavior only for $J_1 \leq 0.7$ and for $J_2 \geq 0.5$. It is also quite noticeable that as we increase the value of $J_1$ the range of $J_2$ values at which $T_{comp}$ exists decreases. The figure also suggests that the value of $T_{comp}$ is more affected by the variation in $J_1$ than the variation in $J_2$. This is clearly shown in the figure when $J_2$ is fixed at high values such as 2 and 2.5, and the value of $J_1$ increases, which results in a significant increase in the value of $T_{comp}$.

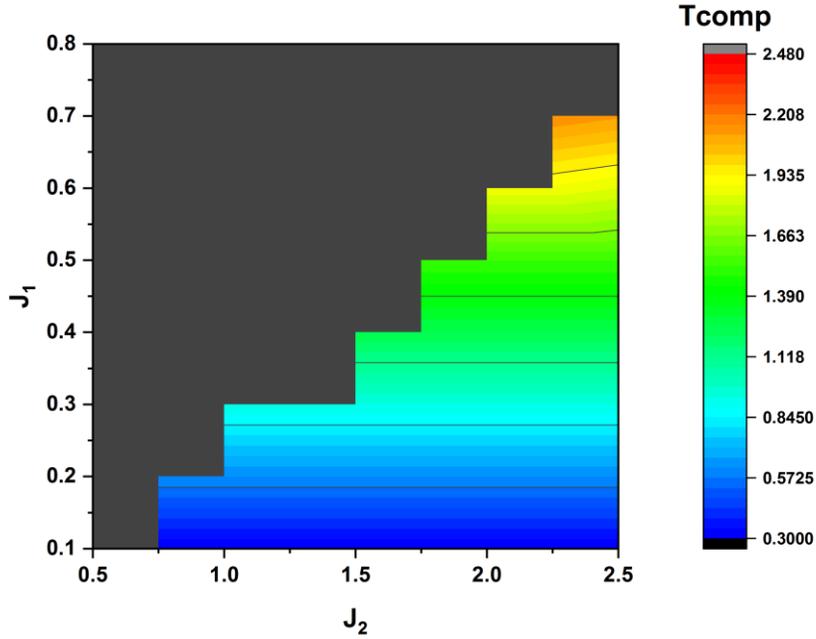

**Fig. 2.** A contour plot of the compensation temperature of the triple layers cubic structure at different values of the coupling constants $J_1$ and $J_2$ at fixed parameters $J_{12} = -0.3$ and $D = 0$.

To investigate the effect of the crystal field $D$ on the system critical and compensation temperatures, the crystal field values will be varied in the simulations at different fixed values of $J_1$ and $J_2$ at which the system compensation temperature is present and absent, as shown in Fig. 2. For all simulations, the coupling constant $J_{12}$ was fixed at -0.3, and the external magnetic field $B$ was fixed at zero.

We start by fixing the parameters $J_1 = 0.5$ and $J_2 = 2$, at which the system exhibits a compensation temperature, as shown in Fig. 2, and varying the crystal field value from -2 to 4 in 0.2 steps. The influence of $D$ on the system $T_c$ and $T_{comp}$ is shown in Fig. 3. For $D = 0$ the system exhibits a compensation temperature, which is confirmed by Fig. 2. For $D < 0$, the system compensation temperature decreases as the crystal field decreases until it disappears below the threshold value $D = -1.4$, below which no compensation exists. For $D > 0$, the system compensation temperature increases as the crystal field increases until it disappears above the threshold value $D = 2.8$; above this value, the system does not possess any compensation behavior. For the critical temperature, its value remained fixed for all values of the crystal field.

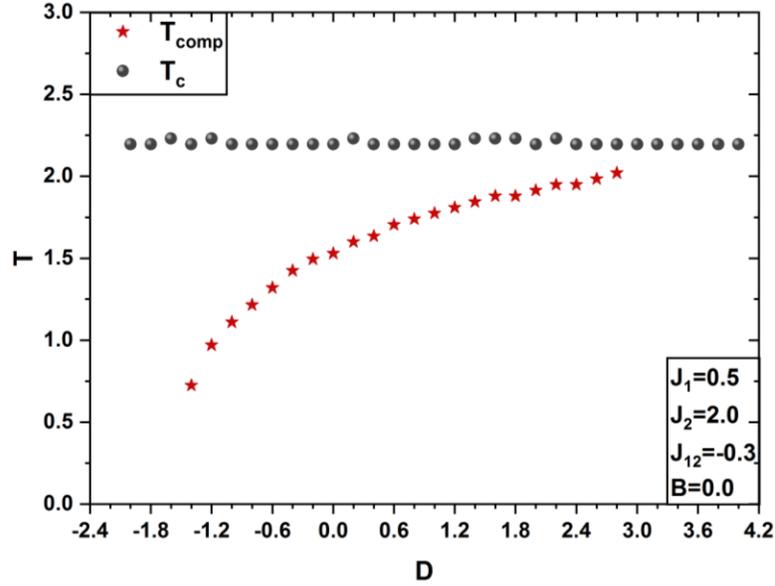

**Fig. 3.** The phase diagram of the system in $(T, D)$ plane for $J_1 = 0.5, J_2 = 2, J_{12} = -0.3$, and $B = 0$.

In this case, our results are confirmed by the sublattice magnetization, total magnetization, susceptibility, and specific heat, as shown in Fig. 4. Here, we fix $J_1 = 0.5, J_2 = 2, J_{12} = -0.3$ and $B = 0$ for different values of the crystal field $D = -1.6, -1.4, -1, 0, 1, 2$.

As shown in Fig. 4(b), for negative values of the crystal field $D = -1, -1.4$, the magnetization curve exhibits two zero points corresponding to the compensation and critical temperatures. The position of the first crossing point for $D = -1.4$ is shifted to the left relative to the first crossing point for $D = -1$. This confirms that for negative values of $D$, the system compensation temperature decreases with decreasing $D$. For $D = -1.6$, which is below the threshold value of $D = -1.4$, the magnetization curve exhibits just one zero point, which corresponds to the critical temperature. This confirms that below the threshold value of $D = -1.4$ the system does not possess a compensation temperature. For $D = 0, 1, 2$, the magnetization curves have two zero points corresponding to the compensation and critical temperatures, respectively. It is clear from the figure that as we increase the value of $D$, the first crossing point of the magnetization curve shifts to the right. This confirms that for positive $D$ values, increasing the value of the crystal field yields a significant increase in the system compensation temperature. For all $D$ values, the second crossing point of the magnetization curve remained fixed. This confirms that the critical temperature of the system remains unchanged with the changing crystal field. It is also noticeable that the magnetization curves have different shape types, as mentioned in Neel's theory [35]. For $D = -1.6$, the $Q$–type magnetization curve is clearly observed, and for the remaining $D$ values, the $N$–type magnetization curves are clearly observed.

As shown in Fig. 4(a), sublattices A and B start from a fully ordered state at low temperature with magnetization values of 1 and -0.5. As the temperature increases, the disorder in the sublattices increases, and the magnetization changes accordingly. It is also obvious from the figure that the sublattice magnetization curves satisfy Eq. (9) and Eq. (10) for all $D$ values, except for $D = -1.6$. This further confirms that the system does not possess compensation behavior below the threshold value of $D = -1.4$. It is also apparent that the magnetization of sublattice A approaches zero for large negative values of $D$. The system tends to prefer the zero state (S=0) because for such large values of D, the energy of the system increases, as shown in Fig. 4c.

The total specific heat and susceptibility of the system shown in Fig. 4(c) and Fig. 4(d) have two peaks for $D \geq -1.4$, corresponding to the compensation and critical temperature, respectively. While the position of the second peak is fixed for all $D$ values, the position of the first shifts to the right with increasing the value of $D$. This further confirms that the critical temperature remains fixed for all $D$ values, and the compensation temperature increases with increasing the value of $D$. For $D = -1.6$ which is below the threshold value, the total specific heat and susceptibility curves exhibit one peak at the critical temperature. This further confirms that the system does not have compensation behavior for $D < -1.4$.

To further investigate the effect of the crystal field $D$ on the system critical and compensation temperatures, we varied the value of $D$ for fixed $J_1$ and $J_2$ at which the system does not possess a compensation temperature. The main goal of this calculation is to investigate whether the crystal field can cause the compensation temperature to exist in the system.

We start by fixing $J_1 = 0.6$ and $J_2 = 1.5$ at which the system compensation temperature is absent, as shown in Fig. 2, and varying the value of $D$ from -2 to 3 in 0.2 steps. The crystal field effect on the system critical and compensation temperatures is presented in the phase diagram shown in Fig. 5. As shown in the figure, the critical temperature of the system remains fixed in the range $-2 \leq D \leq -0.8$, and for $D > -0.8$ the value of $T_c$ increases linearly with increasing $D$. For the compensation temperature, the figure shows that the system does not possess a value of $T_{comp}$ for $D = 0$, which is consistent with the results presented in Fig. 2. For $D < 0$, we observe that the system exhibits a compensation temperature only in the range $-1.6 \leq D \leq -1.0$ and the value of $T_{comp}$ increases with increasing $D$. This indicates that the presence of the crystal field can indeed cause compensation behavior to exist in the system. For $D > 0$, the system does not have a compensation temperature for all $D$ values.

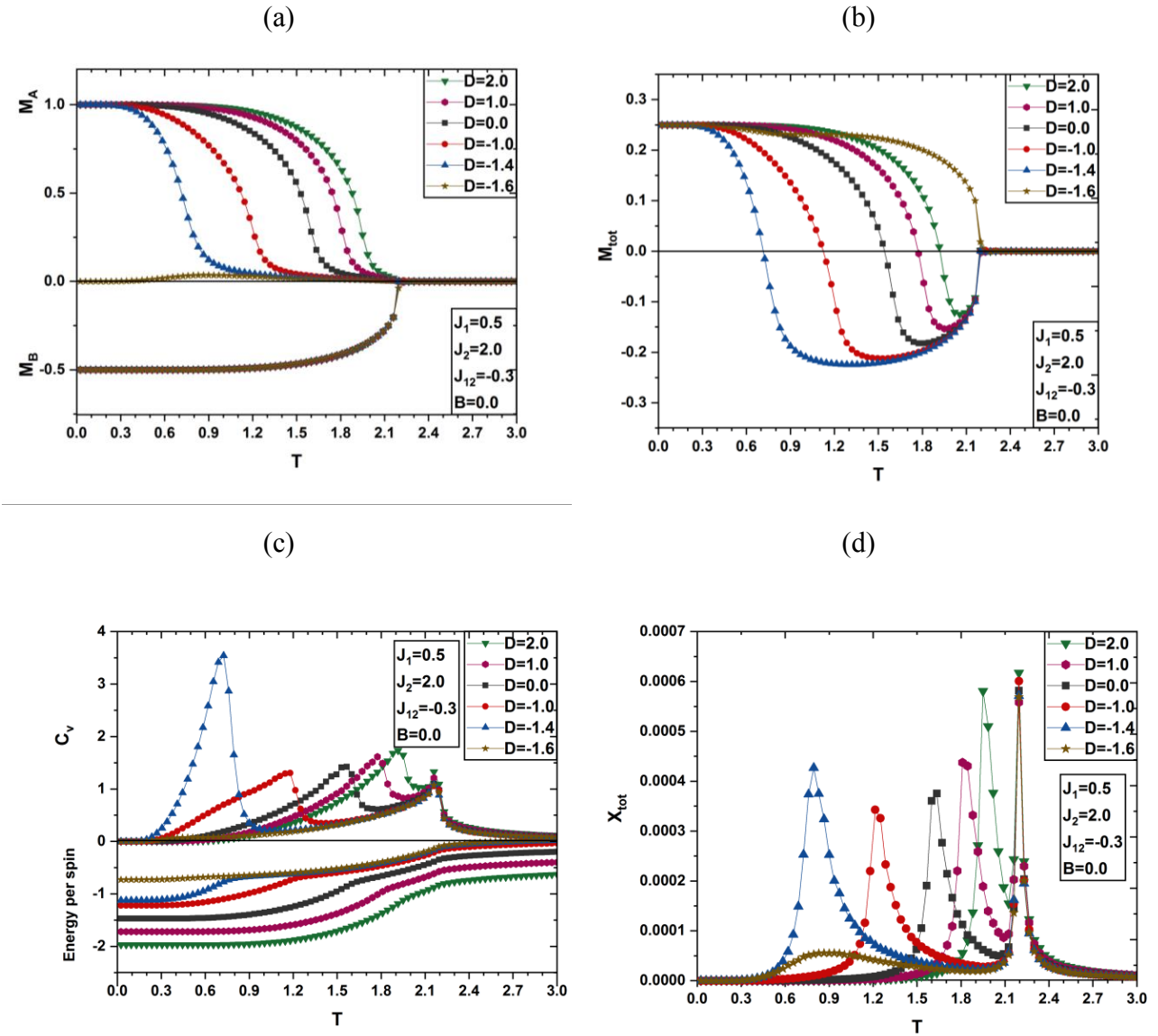

Fig. 4. The temperature dependencies of (a) sublattices magnetization, (b) total magnetization, (c) specific heat and the energy per spin where the peaks in the specific heat curves correspond to the inflection points in the energy per spin curves, (d) total susceptibilities, at $J_1 = 0.5$, $J_2 = 2.0$, $J_{12} = -0.3$, $B = 0$ for different values of the crystal field $D$.

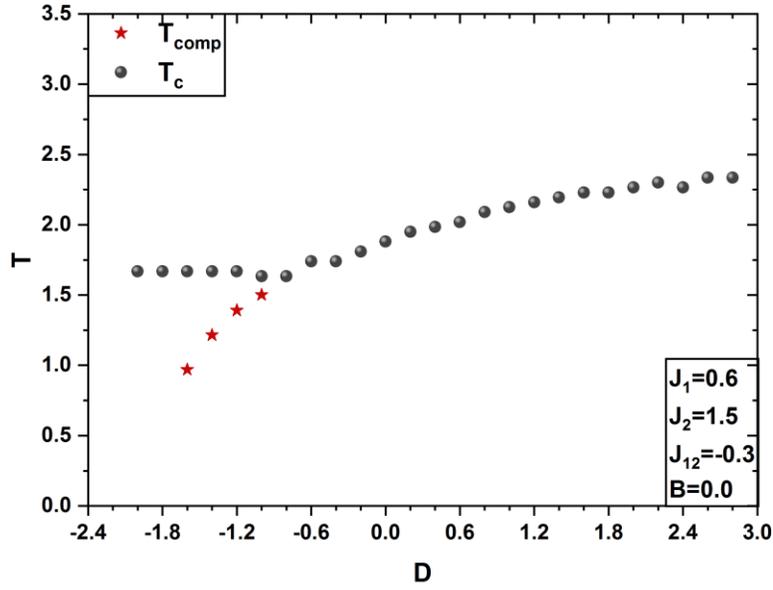

**Fig. 5.** The phase diagram of the system in $(T, D)$ plane at fixed $J_1 = 0.6$, $J_2 = 1.5$, $J_{12} = -0.3$ and $B = 0$.

In this case, our results are confirmed by sublattice magnetization, total magnetization, susceptibility, and specific heat, as shown in Fig. 6. Here we fix $J_1 = 0.6$, $J_2 = 1.5$, $J_{12} = -0.3$ and $B = 0$ for a different value of the crystal field $D = -1.6, -1.4, -1.0, 0, 1.0$.

As shown in Fig. 6(b), for $D = -1.6, -1.4$, and $-1.0$, the total magnetization curve has two zero points at the compensation and critical temperatures. This confirms that the system exhibits a compensation behavior in the range of $-1.6 \leq D \leq -1.0$. While the position of the first peak shifts to the right with increasing $D$, the position of the second peak remains fixed. This is consistent with our observation that $T_{comp}$ increases linearly with $D$ in the range $-1.6 \leq D \leq -1.0$, and $T_c$ remains fixed in the range $-2 \leq D \leq -0.8$. For $D = 0$ and $1$, the magnetization curve has one zero point corresponding to the critical temperature. This ensures that the system only possesses a compensation behavior for $-1.6 \leq D \leq -1.0$. The position of the crossing point for $D = 1$ is shifted to the right relative to the crossing point for $D = 0$. This ensures that $T_c$ increases with increasing $D$ in the range $D > -0.8$. The magnetization curves exhibited different shape types. P-type magnetization curves for $D = 0.0$, and $1.0$, N-type for the rest of the $D$ values.

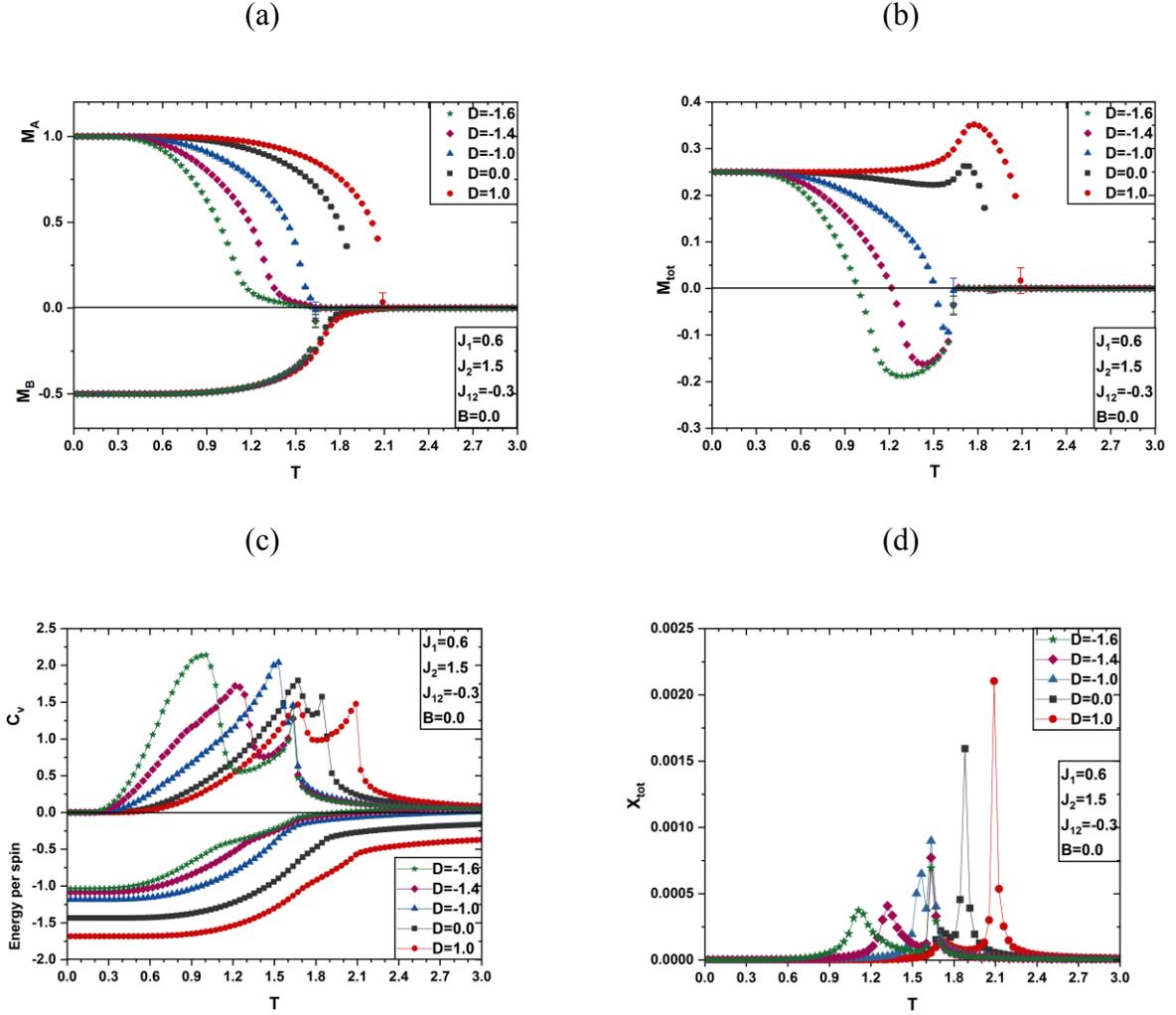

**Fig. 6**. The temperature dependencies of (a) sublattices magnetization, (b) total magnetization, (c) specific heat and the energy per spin where the peaks in the specific heat curves correspond to the inflection points in the energy per spin curves, (d) total susceptibilities, at $J_1 = 0.6$, $J_2 = 1.5$, $J_{12} = -0.3$, $B = 0$ for different values of the crystal field $D$.

As shown in Fig. 6(a), the sublattice magnetization curves satisfy Eq. (9) and (10) only for $D = -1.6, -1.4, -1.0$ resulting in a zero-total magnetization at $T_{comp}$. This further confirms that the system possesses a compensation behavior only in the range $-1.6 \leq D \leq -1.0$.

The total specific heat and susceptibility of the system shown in Fig. 6(c) and 6(d) have two peaks for $D = -1.6, -1.4, -1.0$ corresponding to $T_{comp}$ and $T_c$ respectively. This confirms that $T_{comp}$ exists in the range $-1.6 \leq D \leq -1.0$. The position of the first peak shifts to the right as $D$ increases, and the position of the second peak remains fixed. This is consistent with our analysis that $T_{comp}$ increases linearly with $D$ in the range $-1.6 \leq D \leq -1.0$ and $T_c$ remains fixed in the range $-2 \leq D \leq$

$-0.8$. For $D = 0$ and 1, the total specific heat and susceptibility curves exhibit one peak at the critical temperature. This further confirms that the system possesses a compensation behavior only in the range $-1.6 \leq D \leq -1.0$. The peak position is shifted to the right for $D = 1$ relative to the peak position for $D = 0$. This agrees with our analysis that the value $T_c$ increases with $D$ in the range $D > -0.8$.

To further investigate the case in which the system compensation temperature is absent, we choose other values of $J_1$ and $J_2$ at which the system does not possess compensation behavior. Here, we fix $J_1 = 0.8$ and $J_2 = 2$, and we vary $D$ from -2 to 3 in 0.2 steps. The effect of the crystal field on the system compensation and critical temperatures is presented in the phase diagram shown in Fig. 7.

As shown in the figure, the system $T_c$ value remains fixed in the range $-2 \leq D \leq -1.0$. After the threshold value $D > -1.0$, the value of $T_c$ increases linearly with $D$. For the compensation temperature, the figure shows that $T_{comp}$ exist for negative $D$ values in the range $-2 \leq D \leq -1.2$. This further confirms that the presence of the crystal field leads to a compensation temperature that exists in the system. Within the range $-2 \leq D \leq -1.2$, the value of $T_{comp}$ increases linearly with the crystal field. For $D > -1.2$, the system does not possess any compensation behavior for all $D$ values.

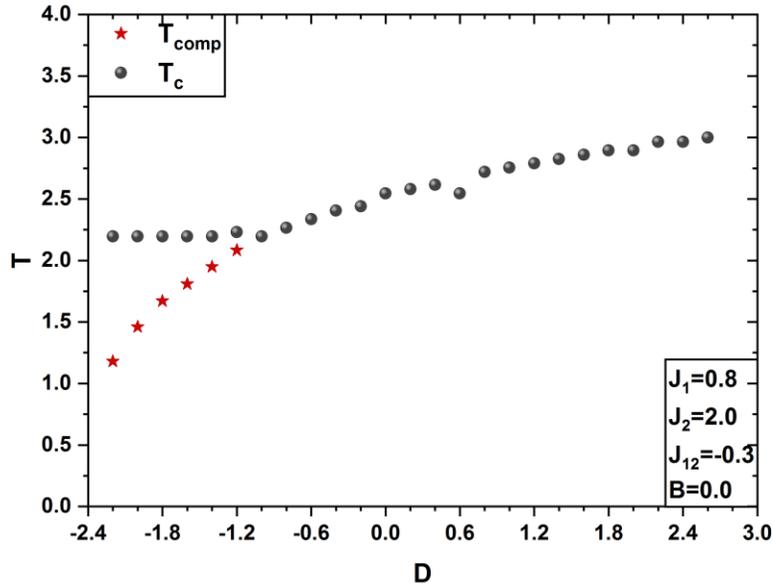

**Fig. 7.** The phase diagram of the system in $(T, D)$ plane at fixed $J_1 = 0.8, J_2 = 2, J_{12} = -0.3$, and $B = 0$.

In this case, our results are confirmed by the total magnetization curves shown in Fig. 8. Here, we fix $J_1 = 0.8$, $J_2 = 2$, $J_{12} = -0.3$ and $B = 0$ for different values of the crystal field $D = -2, -1.2, -1, 0$, and $1.0$.

As shown in the figure, for $D = -1.2, -2$, which is in the range $-2 \leq D \leq -1.2$, the magnetization curve has two zero points corresponding to the compensation and critical temperatures. This confirms that the system possesses compensation behavior in the range $-2 \leq D \leq -1.2$. For $D = -1, 0, 1$, the magnetization curves have one zero point at the critical temperature. This confirms that $T_{comp}$ exists only in the range $-2 \leq D \leq -1.2$. The position of the first crossing point increases with increasing $D$ for $D = -1.2, -2$ this verifies that the value of $T_{comp}$ increases with $D$ in the range $-2 \leq D \leq -1.2$. The position of the crossing point corresponding to the $T_c$ remains fixed for $D = -1, -1.2, -2$, and it shifts to the right with increasing $D$ for $D = 0, 1$. This confirms that $T_c$ remains fixed for $D \leq -1.0$, and it increases with $D$ for $D > -1.0$. In this case, the magnetization curves show P-type magnetization for $D = 0.0, 1.0$, H-type magnetization for $D = -1.0$, and N-type magnetization for $D = -1.2, -2.0$.

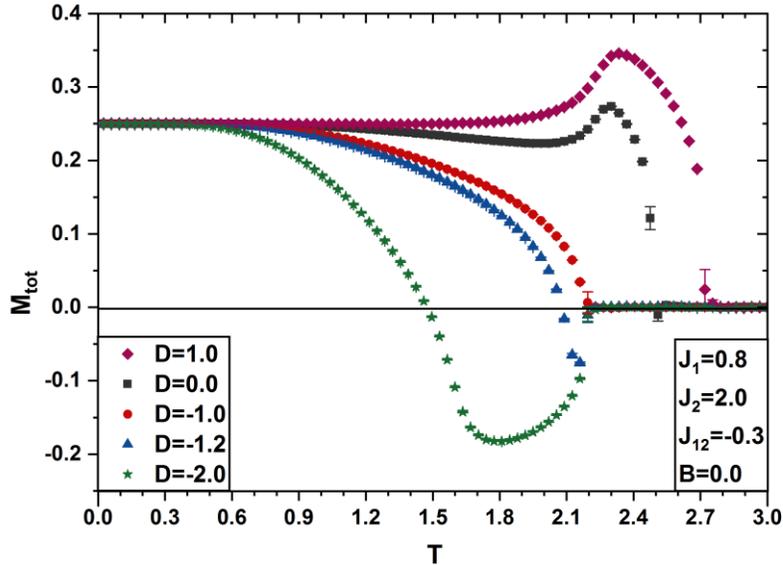

**Fig. 8**. The temperature dependencies of total magnetization at fixed $J_1 = 0.8$, $J_2 = 2$, $J_{12} = -0.3$, $B = 0$ for different values of the crystal field $D$.

Finally, it is worth mentioning that the results obtained in this study are insensitive to the structure size. We performed simulations with different structure sizes, including $L = N = 40, 48, 56, 64, 72,$ but no significant differences were found in the results presented here.

3.2 *Hysteresis loops*

In this section, the magnetic hysteresis loops of the mixed spin (1-1/2-1) three-layer system of the cubic structure shown in Fig. 1 will be studied in detail. The main objective of this study was to investigate whether the system possesses multiple-step hysteresis loops. Furthermore, we aim to investigate the effects of the Hamiltonian parameters in Eq. (1) on those loops.

To investigate the effect of the crystal field $D$ on the system hysteresis loop, we fix the parameters $J_1 = 0.5, J_2 = 2.0, J_{12} = -0.3$ and temperature $T = 1.0$ and varied the value of $D$. For this case, the hysteresis loops for negative values of $D = -2, -1.4, -1$ are shown in Fig. 9, and positive values of $D = 0, 1, 2$ are shown in Fig. 10.

For the negative $D$ values shown in Fig. 9, it is interesting to observe multiple-step hysteresis loops with the variation of $D$. For $D = -2.0$ a single hysteresis loop is observed, and with increasing $D$ to -1.4, the shape of the hysteresis loop changes, but it remains a single loop. When $D$ to -1.0, the shape of the loop changes as well, and a two-step hysteresis loop is observed. It is noticeable that the area of the loops remains unchanged with the variation in $D$ in this case. In addition, one can observe that the shape of sublattice-B loops is insensitive to the variation in $D$; accordingly, the change in the loop shape can be attributed to the change in the sublattice-A loop, which can be observed as well.

As shown in Fig. 10, for positive values of $D = 0, 1, 2$, single hysteresis loops were observed for all $D$ values. The shape of these loops is not sensitive to the variation in $D$. Furthermore, the area of the hysteresis loops is not sensitive to the variation in $D$ in this case. In addition, the shape of the sublattice loops appears to be insensitive to the variation of $D$ as well.

The effect of temperature on the magnetic hysteresis loops was considered by varying the value of $T$ for several fixed values of the Hamiltonian parameters. We start by fixing the parameters $J_1 = 0.6, J_2 = 1.5, J_{12} = -0.3$ and $D = -1.4$, and we varied the temperature to the values $T = 1.0, 1.2, 1.4$. The magnetic hysteresis loops in this case are shown in Fig. 11. A multiple-step hysteresis loop was observed with the variation in $T$. It is interesting to observe that the hysteresis loop changes from a single loop at $T = 1.0$ to a two-step loop at the compensation temperature $T = 1.2$ and then returns to a single loop at $T = 1.4$. In addition, it is quite noticeable that the area of the loop decreases with increasing temperature. The increase in $T$ seems to reduce sublattice A and B loop areas, causing the total loop area to decrease.

It is interesting to investigate the hysteresis loops for high-temperature values, as shown in Fig. 12. Here, we fix the parameters $J_1 = 0.6, J_2 = 1.5, J_{12} = -0.3$ and $D = 0.0$, and we fix the temperature at a high value of $T = 1.8$. As shown in the figure, the hysteresis loop exhibits paramagnetic behavior at high temperatures.

To further investigate the effect of temperature on the hysteresis loops, the temperature was varied at different fixed values of the Hamiltonian parameters. In Fig. 13, the hysteresis loops are presented at fixed $J_1 = 0.8, J_2 = 2.0, J_{12} = -0.3$ and $D = -1.4$ for various temperature values of $T = 1.7, 1.9, 2.1$. In Fig. 14, the hysteresis loops are presented at fixed $J_1 = 0.8, J_2 = 2.0, J_{12} = -0.3$ and $D = -1.0$ for various temperature values of $T = 1.5, 2.35$. Finally, in Fig. 15, the hysteresis loops are presented at fixed $J_1 = 0.8, J_2 = 2.0, J_{12} = -0.3$ and $D = 1.0$ for a high-temperature value of $T = 2.1$.

As shown in Fig. 13, the hysteresis loop shape changes from a single at $T = 1.7$ to a two-step loop at the compensation temperature $T = 1.9$ and returns to a single loop at $T = 2.1$. The area of the sublattice loop shrinks as the temperature increases, causing the area of the total magnetization to shrink. It is quite noticeable that the hysteresis loop shows a paramagnetic behavior at high-temperature $T = 2.1$ in this case.

As shown in Fig. 14, a single hysteresis loop was observed for both temperatures, $T = 1.5, 2.35$. The area of the loop decreases as the temperature increases, and it turns into a paramagnetic state at a high-temperature $T = 2.35$. The hysteresis loops presented in Fig. 15 show paramagnetic behavior for a fixed high-temperature $T = 2.1$.

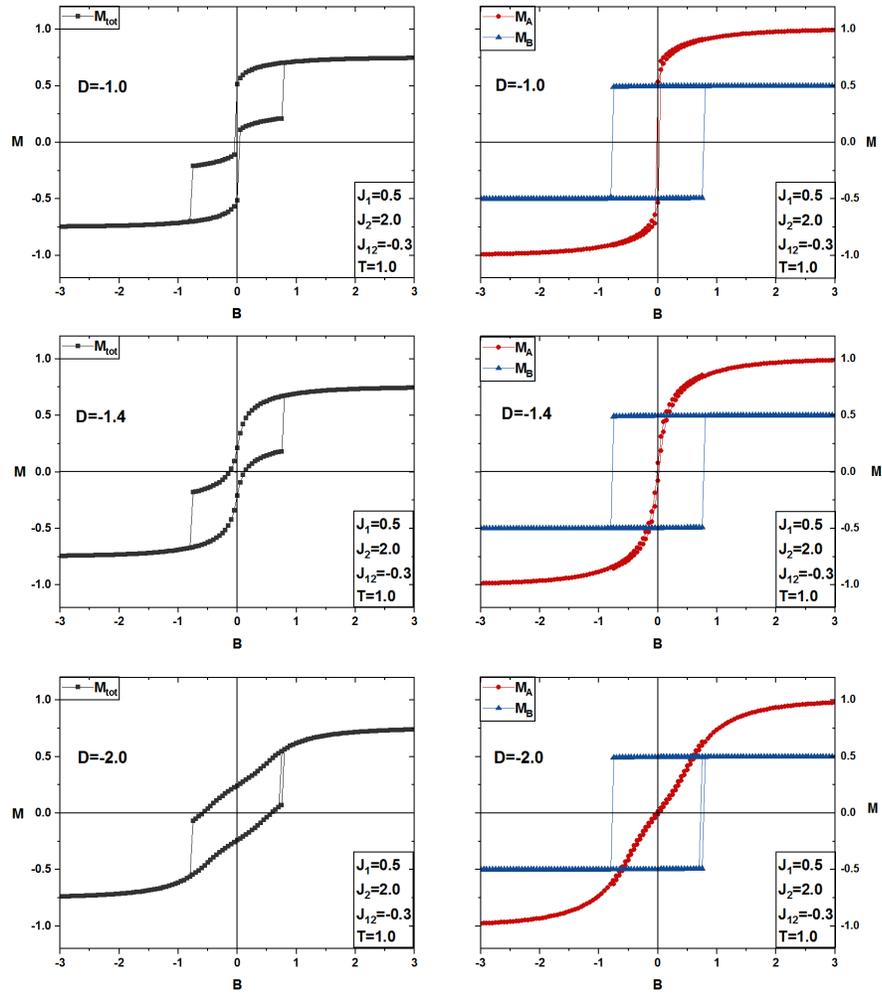

**Fig. 9.** The magnetic hysteresis cycles for different crystal field values $D$ ($D = -2.0, -1.4, -1.0$) for (a) total magnetization (b) the A and B sublattices magnetizations at fixed parameters $J_1 = 0.5, J_2 = 2.0, J_{12} = -0.3, T = 1.0$.

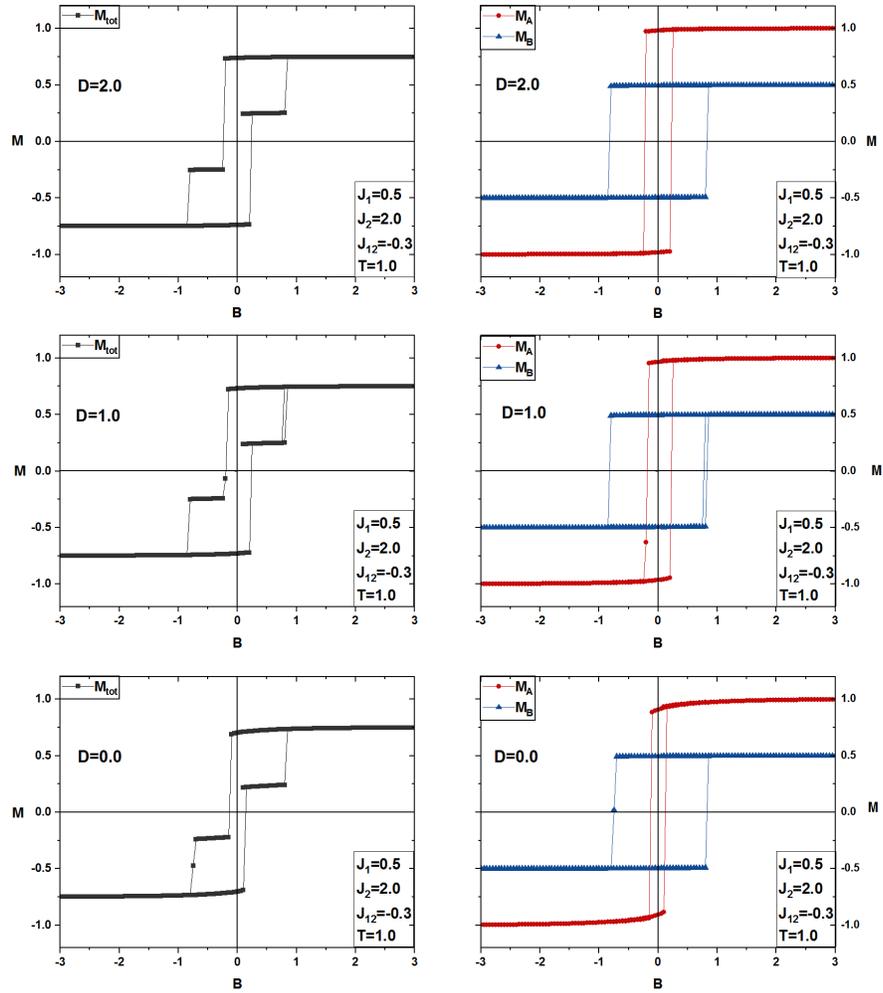

**Fig. 10.** The magnetic hysteresis cycles for different crystal field values $D$ ($D = 0, 1, 2$) for (a) total magnetization (b) the A and B sublattices magnetizations at fixed parameters $J_1 = 0.5, J_2 = 2.0, J_{12} = -0.3, T = 1.0$.

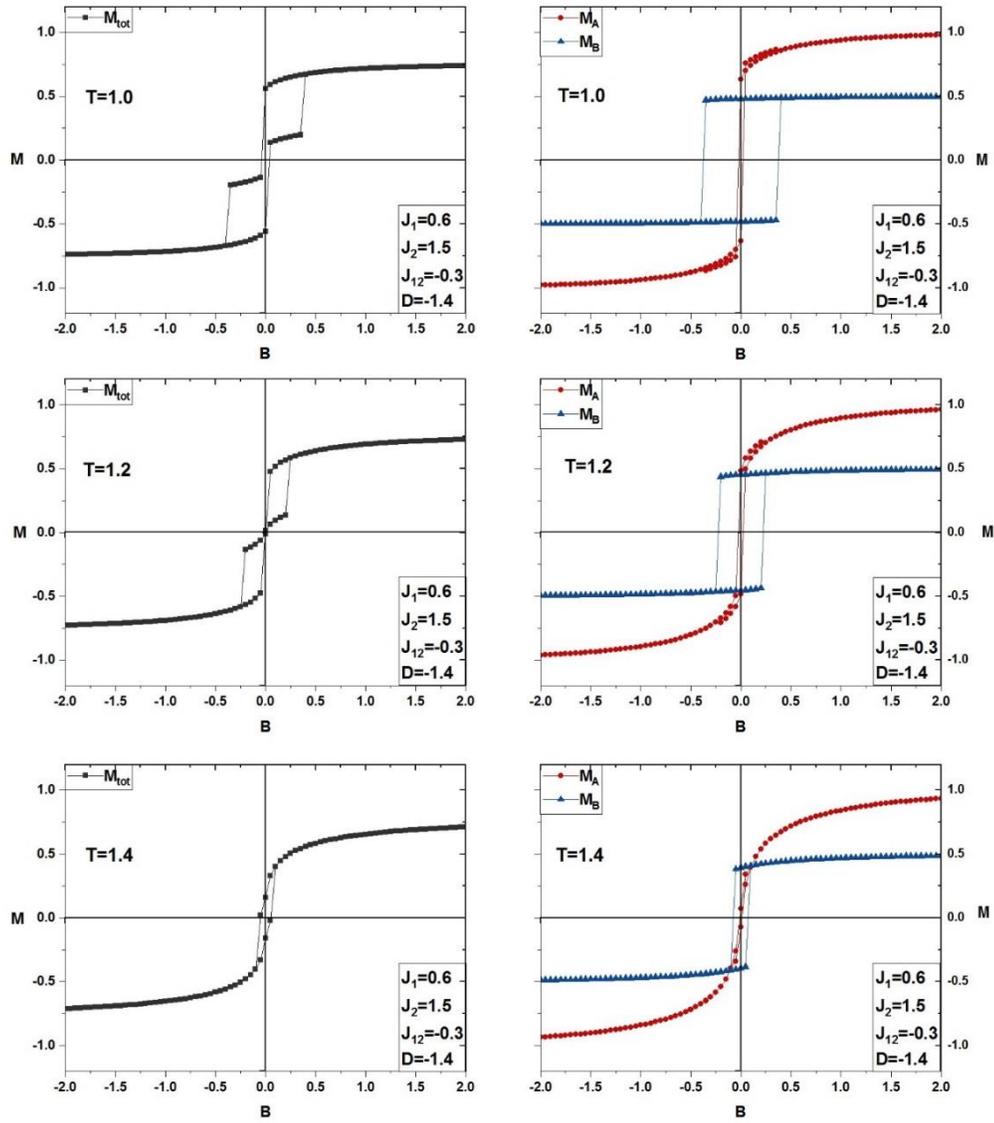

**Fig. 11.** The magnetic hysteresis cycles for different temperature values $T$ ($T = 1, 1.2, 1.4$) for (a) total magnetization (b) the A and B sublattices magnetizations at fixed parameters $J_1 = 0.6, J_2 = 1.5, J_{12} = -0.3$ and $D = -1.4$.

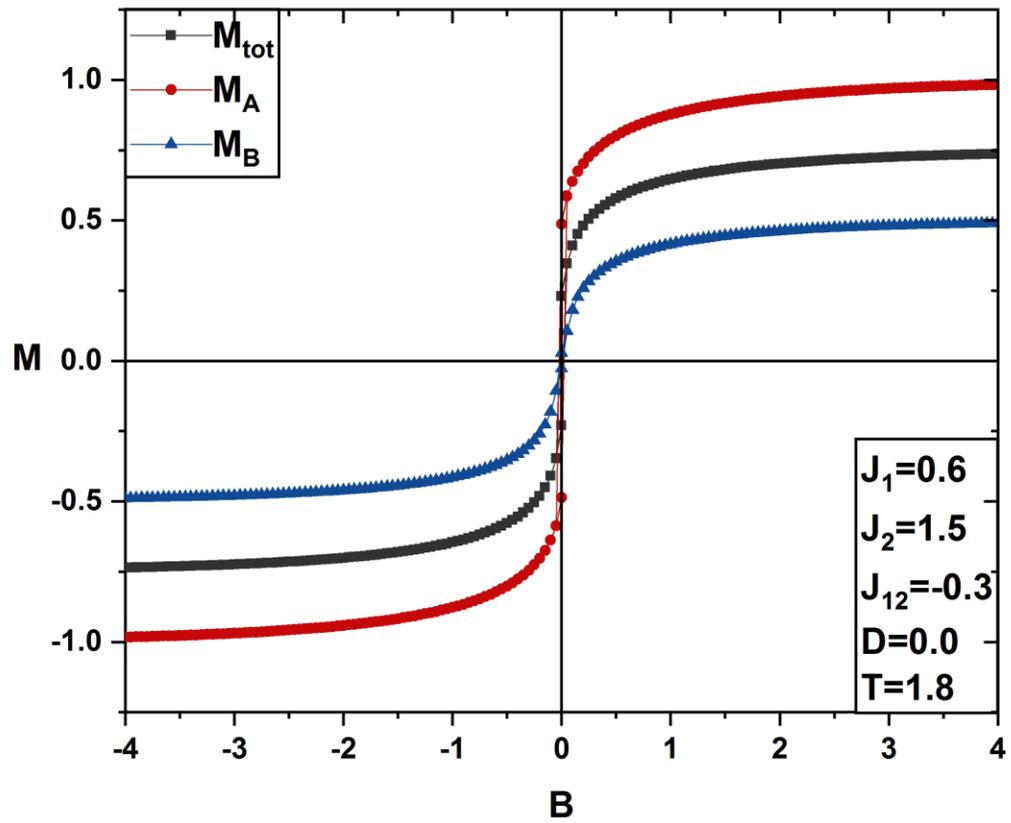

**Fig. 12.** The magnetic hysteresis cycles for total magnetization and sublattices magnetizations at fixed parameters $J_1 = 0.6, J_2 = 1.5, J_{12} = -0.3, D = 0.0, T = 1.8$.

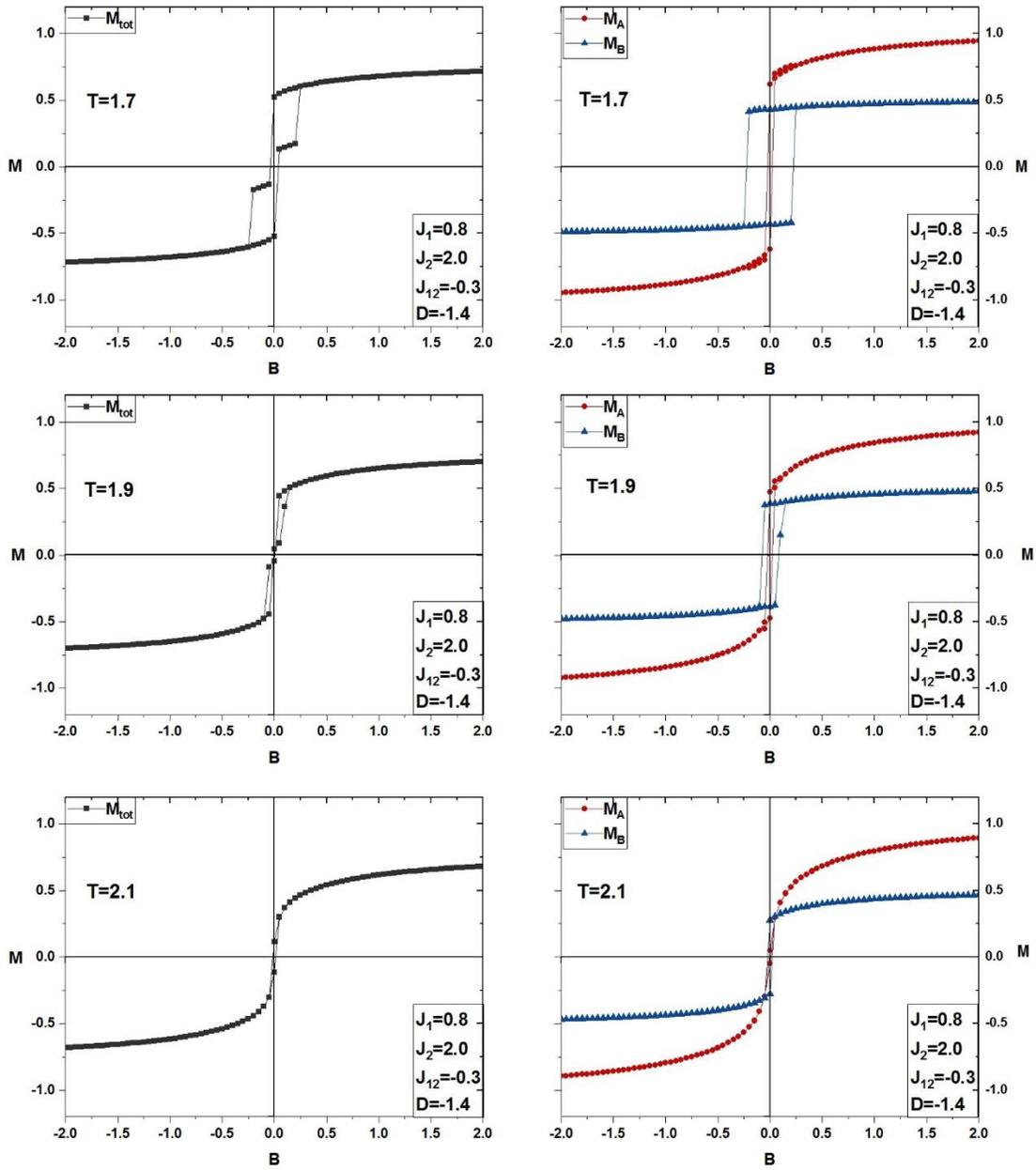

**Fig. 13.** The magnetic hysteresis cycles for different temperature values $T$ ($T = 1.7, 1.9, 2.1$) for (a) total magnetization (b) the A and B sublattices magnetizations at fixed parameters $J_1 = 0.8, J_2 = 2.0, J_{12} = -0.3$ and $D = -1.4$.

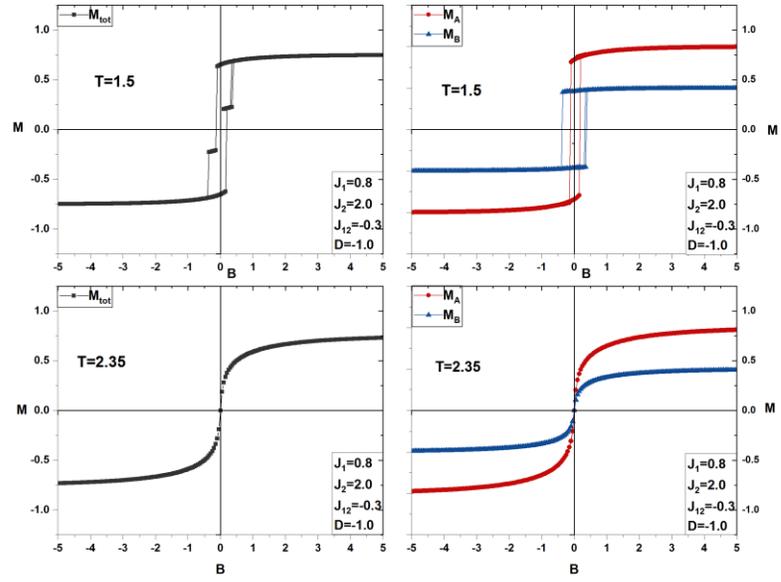

**Fig. 14.** The magnetic hysteresis cycles for different temperature values $T$ ($T = 1.5, 2.35$) for (a) total magnetization (b) the A and B sublattices magnetizations at fixed parameters $J_1 = 0.8, J_2 = 2.0, J_{12} = -0.3$ and $D = -1.0$.

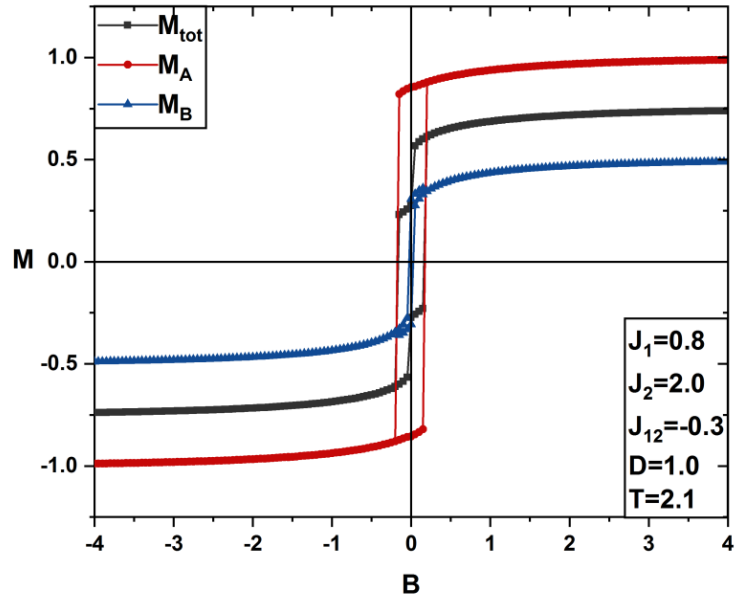

**Fig. 15.** The magnetic hysteresis cycles for total magnetization and sublattices magnetizations at fixed parameters $J_1 = 0.8, J_2 = 2.0, J_{12} = -0.3, D = 1.0, T = 2.1$.

## 4. Conclusions

The impact of the crystal field on the system critical and compensation temperatures and the magnetic and thermodynamic properties of the system were explored using Monte Carlo simulations. The system $T_{comp}$ was found to be sensitive to exchange and coupling constants $J_1$ and $J_2$. A well-defined value of $T_{comp}$ was found to exist only in the range $J_1 \leq 0.7$ and for $J_2 \geq 0.5$. The range of $J_2$ values at which $T_{comp}$ exist was found to narrow as the value of $J_1$ increase.

The crystal field was found to have a profound impact on $T_{comp}$ and $T_c$. For the $J_1$ and $J_2$ values at which the system possesses a compensation temperature, $T_{comp}$ was found to exist between two threshold values of $D$. The value of $T_{comp}$ was found to increase linearly with $D$, and $T_c$ remains fixed for all $D$ values. For the $J_1$ and $J_2$ values at which the system $T_{comp}$ is absent, the presence of the crystal field was found to cause $T_{comp}$ to exist in the system in a short range of negative $D$ values. In this range, $T_{comp}$ was found to increase linearly with $D$ and $T_c$ remains fixed. Beyond this range, the compensation temperature was absent, and $T_c$ increases linearly with $D$.

The magnetization and energy curves were observed to be continuous, which confirms that the system does not show any first-order phase transition even for large values of $D$. A two-step hysteresis loop was observed for several negative $D$ values. No obvious change in the area of the loop was observed with the variation of $D$. On the other hand, the temperature variation was found to significantly impact the shape and area of the loops. The loop area was found to decrease with increasing temperature, and two-step hysteresis loops were observed at different compensation temperatures. In addition, the paramagnetic behavior was observed at different temperatures. Our results regarding the temperature effect on the hysteresis loops are in good agreement with those found for the rectangular Ising nanoribbon [36], Ising ladder-like boronene nanoribbons [37], Ising graphene nanoribbons [38], cylindrical Ising nanowires [11], and borophene structures [39].

**Acknowledgment**

This work was supported by the Jordan University of Science and Technology under Grant (20200039) and Khalifa University under grant CIRA-2019-043.

**Conflict of interest statement**

The authors declare that no competing interests exist

## Data access statement

All relevant data are within the paper and its Supporting Information files.